\documentclass{ws-procs9x6}

\begin{document}
 {\centerline {14th Conference on Waves and Stability in Continuous
Media, Baia Samuele, Sicily, Italy, 30 June - 7 July, 2007}}
{\centerline {Edited by N Manganaro, R Monaco   and S Rionero,
 World Scientific 2008,   pp. 327-326, ISBN 978-981-277-234-3.}}

\title{A MECHANICAL MODEL FOR LIQUID NANOLAYERS}

\author{H. GOUIN}

\address{C.N.R.S.  U.M.R. 6181 \& Universit\'e d' Aix-Marseille  \\ Case 322, Av. Escadrille
 Normandie-Niemen, 13397 Marseille Cedex 20 France\\
E-mail: henri.gouin@univ-cezanne.fr}

\begin{abstract}
Liquids in contact with solids are submitted to intermolecular
forces  making liquids   heterogeneous and   stress tensors  are not
any more spherical as in homogeneous bulks.
  The aim  of this article is to show that a square-gradient functional representing
liquid-vapor interface free energy  corrected with a liquid density
functional at  solid surfaces is a well adapted model to study
structures of very thin nanofilms near solid walls. This result
makes it possible to study the motions of liquids in nanolayers and
to generalize the approximation of lubrication in long wave
hypothesis.
\end{abstract}


\section{Introduction}
At the end of the nineteenth century, the fluid inhomogeneity in
liquid-vapor interfaces was taken into account by considering a
volume energy depending on space density derivative \cite{Widom}.
This van der Waals square-gradient functional  is unable to model
repulsive force contributions and misses the dominant damped
oscillatory packing structure of liquid interlayers near a substrate
wall \cite{chernov1}. Furthermore, the decay lengths are
correct only close to the liquid-vapor critical point where the damped
oscillatory structure is subdominant
 \cite{Evans1}. In mean field theory, weighted density-functional
  has been used to explicitly demonstrate the dominance
 of this structural contribution in van der Waals thin films and to take into
  account long-wavelength capillary-wave fluctuations as in papers that
 renormalize the square-gradient functional to include capillary wave fluctuations \cite{Fisher}.
 In contrast, fluctuations strongly damp
 oscillatory structure and it is mainly for this reason that van der Waals'
 original prediction of a \emph{hyperbolic tangent}
 is so close to simulations and experiments
 \cite{rowlinson}.
 The recent development of experimental technics allows us to observe
physical phenomena at length scales of a few nanometers
\cite{Bhushan}. To get an  analytic expression in
density-functional theory for liquid
 film of a few nanometer thickness near a solid wall, we add a liquid density-functional
 at the solid    surface to the square-gradient functional  representing closely
 liquid-vapor interface free energy. This kind of functional
 is well-known in the literature \cite{Fisher1}.
 It was used by Cahn in a phenomenological form,
 in a well-known paper studying wetting near a critical point \cite{Cahn0}.
 An asymptotic expression is obtained in \cite{gouin}
 with an approximation of hard sphere molecules and London potentials for
  liquid-liquid and solid-liquid interactions:   we took into account the power-law
  behavior which is dominant in a thin liquid film in contact with a solid.\\
  For fluids submitted to this density-functional, we recall the equation of   motion
  and boundary conditions.
  We point out the definition of disjoining pressure and
 analyze
the consequences of the model. Finally, we study the motions in
liquid nanolayers; these motions are always object of many debates.
Within lubrication and long wave  approximations,   a relation
between disjoining pressure, viscosity of liquid, nanolayer
thickness variations along the layer and tangential velocity  of the
liquid is deduced.

\section{The  density-functional}

The free energy density-functional of an inhomogeneous fluid in a
domain $O$ of boundary $\partial O$  is taken in the form
\begin{equation}
F = \int\int\int_O \varepsilon\ dv + \int\int_{\partial O} \varphi\ ds
.\label{density functional}
\end{equation}
 The first integral is associated with square-gradient approximation when we
  introduce a specific
free energy of the fluid at a given temperature $\theta$,
$
\varepsilon =\varepsilon(\rho,\beta)
$
as a function of density $\rho$ and  $\beta= (\rm grad\, \rho)^2$.
Specific free energy $\varepsilon$ characterizes together fluid
properties of \emph{compressibility} and \emph{molecular
capillarity} of liquid-vapor interfaces.
 In accordance  with gas kinetic
theory, $\lambda =2\rho\, \varepsilon _{\beta }^{\prime }(\rho,
\beta)$ is assumed  to be constant  at given temperature \cite{Rocard} and
\begin{equation}
\rho \,\varepsilon =\rho \,\alpha (\rho)+\frac{\lambda
}{2}\,(\text{grad\ }\rho )^{2},  \label{internal energy}
\end{equation}
where term $ ({\lambda }/{2})\,(%
\mathrm{grad\ \rho )^{2}}$  is added  to the volume free energy
$\rho \,\alpha (\rho)$ of a compressible fluid. Specific free energy
$\alpha $ enables to connect  continuously  liquid and vapor bulks
and
 pressure $P(\rho)=\rho
^{2}\alpha _{\rho }^{\prime }(\rho )$ is similar to van der Waals
one. Near a solid wall, London potentials of liquid-liquid and
liquid-solid interactions are
\begin{equation*}
\left\{
\begin{array}{c}
\displaystyle\;\;\;\;\;\;\varphi
_{ll}=-\frac{c_{ll}}{r^{6}}\;,\text{ \ when\ }r>\sigma
_{l}\;\;\text{and }\;\ \varphi _{ll}=\infty \text{ \ when \ }r\leq
\sigma _{l}\, ,\  \\
\displaystyle\;\;\;\;\;\;\varphi
_{ls}=-\frac{c_{ls}}{r^{6}}\;,\text{ \ when\ }r>\delta \;\;\text{and
}\;\ \varphi _{ls}=\infty \text{ \ when \ }r\leq \delta \;, \
\end{array}
\right.
\end{equation*}
where $c_{ll}$ and $c_{ls}$ are two positive constants associated
with Hamaker constants, $\sigma _{l}$ and $\sigma _{s}$ denote fluid
and solid  molecular diameters, $\delta =\frac{1}{2}($ $\sigma
_{l}+$ $\sigma _{s})$ is the minimal distance between centers of
fluid and solid molecules \cite{Israel}. Forces between liquid and
solid have short range and can be described simply by adding a
special energy at the surface. This is not  the entire interfacial
energy: another contribution comes from the distortions in the
density profile near the wall \cite{degennes,gouin}.
 For a plane solid wall (at a molecular scale), this
  surface free energy is
\begin{equation}
\phi(\rho)=-\gamma _{1}\rho+\frac{1}{2}\,\gamma_{2}\,\rho^{2}.
\label{surface energy}
\end{equation}
Here $\rho$ denotes the fluid density value  at the wall; constants
$\gamma _{1}$, $\gamma _{2}$  are positive and given by relations
$\displaystyle
 \gamma
_{1}=\frac{\pi c_{ls}}{12\delta ^{2}m_{l}m_{s}}\;\rho _{sol},\
 \gamma _{2}=\frac{\pi c_{ll}}{12\delta^2 m_{l}^{2}}$\,, where $m_{l}$ and $m_{s}$ denote
respectively masses of fluid and solid molecules, $\rho _{sol}$ is
the solid density \cite{gouin}. Moreover, we have $\displaystyle
\lambda = \frac{2\pi
c_{ll}}{3\sigma_l \,m_{l}^{2}}$\,.\\
We consider a horizontal plane liquid interlayer contiguous to its
vapor bulk and in contact with a plane solid wall $(S)$; the z-axis
is perpendicular to the solid surface.
 The liquid film thickness is denoted by $h$.
Conditions in   vapor bulk yield  $\displaystyle {\rm grad}\, \rho
=0$ and
 $\Delta \rho = 0$. Another way to take into account the vapor bulk contiguous to the
liquid interlayer  is to compute a  density-functional of the
complete \emph{liquid-vapor interlayer} by adding a supplementary
surface energy $\psi$ on a geometrical surface $(\Sigma)$ at  $z=h $
to    volume energy (\ref{internal energy}) in liquid interlayer
$(L)$ and surface energy (\ref{surface energy}) on solid wall $(S)$
 \cite{Gavrilyuk}.
 This assumption corresponds to a liquid interlayer included between $z= 0$ and $z=h$,
 a liquid-vapor interface of a few Angstr\"{o}m thickness assimilated to surface $z=h$
 and a vapor layer included between $z=h$ and
$z=\infty$. Due to small   vapor density, let us denote by $\psi$
the surface free energy of a liquid in contact with a vacuum,
\begin{equation}
\psi (\rho )=\frac{\gamma _{4}}{2}\ \rho ^{2}  \label{cl2}
\end{equation}
where   $ \gamma_4\simeq \gamma_2 $ and $\rho$ is the liquid density
in a convenient point inside the liquid-vapor interface
\cite{Gavrilyuk}. Density-functional (\ref{density functional}) of
the liquid-vapor layer gets the final form
\begin{equation*}
F = \int\int\int_{(L)} \varepsilon\ dv + \int\int_{(S)} \phi\ ds +
\int\int_{(\Sigma)} \psi\ ds \label{density functional2}
\end{equation*}

\section{Equation of motion and boundary conditions}

In case of equilibrium,   functional  $F$ is minimal and  yields the
 equation of equilibrium and boundary conditions. In case of motions
we simply add the inertial forces $\rho\, \mathbf{\Gamma}$
  and the dissipative stresses to the results
\cite{gouin4,Gouin1,Gouin05}.
\subsection{Equation of motion}
 The equation of  motion  is  \cite{gouin4,Gouin1}
\begin{equation}
\rho \ \mathbf{\Gamma }=
\text{div}\left(\mathbf{\sigma}+\mathbf{\sigma}_{v}\right) -\rho\;
\text{grad }\Omega  \ , \label{motion0}
\end{equation}
where $ \mathbf{\Gamma}$ is the acceleration, $\Omega $  the body
force potential and $\mathbf{\sigma }$ the stress tensor
generalization
\begin{equation*}
\mathbf{\ \sigma =}-p\,\mathbf{1}-\lambda \;\text{grad\ }\rho \
\otimes \ \text{grad }\rho ,  \label{contrainte}
\end{equation*}
with   $p=\rho ^{2}\varepsilon _{\rho }^{\prime }-\rho \text{ div\textrm{\ }}%
(\lambda \text{ grad }\rho )$. The viscous stress tensor is $\mathbf{
\sigma }_{v} =\kappa_{1}(\text{ tr } {D})\, \textbf{1} +2\,\kappa
_{2}\;{D} $ where  ${D}$ denotes the velocity strain tensor; $\kappa
_{1}$ and $\kappa _{2}$\ are the coefficients of viscosity.\newline
For a horizontal layer, in an orthogonal system of coordinates such
that the third coordinate is the vertical direction, the stress
tensor $\mathbf{\sigma }$ of the thin film takes the form :
\begin{equation*}
\mathbf{\sigma}= \left[
\begin{array}{ccc}
a_1, & 0, & 0 \\
0, & a_2, & 0 \\
0, & 0, & a_3
\end{array}
\right],\quad {\rm with} \quad \left\{
\begin{array}{ccc}
a_1=a_2 = \displaystyle -P+\frac{\lambda }{2}\left(\frac{d\rho
}{dz}\right)^2+\lambda\, \rho\,
\frac{d^2\rho}{dz^2} \\
a_3= \displaystyle -P-\frac{\lambda }{2}\left(\frac{d\rho
}{dz}\right)^2+\lambda\, \rho\, \frac{d^2\rho}{dz^2}
\end{array}\right.
\end{equation*}
Let us consider a thin film of liquid at equilibrium (\emph{gravity
forces are neglected} but the variable of  position is the ascendant
vertical). The equation of equilibrium is :
\begin{equation}
\text{div }\mathbf{\sigma =0}  \label{equilibrium1a}
\end{equation}
Eq. (\ref{equilibrium1a}) yields a constant value for the eigenvalue
$a_3$,
\begin{equation*}
P+\frac{\lambda }{2}\left(\frac{d\rho }{dz}\right)^2-\lambda \rho\,
\frac{d^2\rho}{dz^2}=P_{v_b}\text{.}  \label{equilibrium1b}
\end{equation*}
where $P_{v_{b}}$ denotes  pressure $P(\rho_{v_{b}})$ in the vapor
bulk of density $\rho_{v_{b}}$ bounding the liquid layer.
Eigenvalues $a_1, a_2$ are not constant  but depend on the distance
$z$ to the solid wall \cite{Derjaguin}.  At equilibrium, Eq.
(\ref{motion0}) yields \cite{gouin4}:
\begin{equation}
\text{grad }\left[ \ \mu \left(\rho\right) -\lambda \Delta \rho\
\right] =0\mathrm{,}  \label{equilibrium2a}
\end{equation}
where $\mu $ is the chemical potential at temperature $\theta $
defined to an unknown additive constant. The chemical potential is a
function of $P$ (and $\theta$)  but it can be also
expressed as a function of $\rho$ (and $\theta$).  We choose as
\emph{reference chemical potential} $\mu _{o}=\mu _{o}(\rho)$   null
for bulks of densities $\rho _{l}$ and $\rho _{v}$ of phase
equilibrium. Due to Maxwell
 rule,
the volume free energy associated with $\mu _{o}$ is
$g_{o}(\rho)-P_{o}$  where   $ P_{o}=P(\rho _{l})=$ $P(\rho _{v})$
is the bulk pressure  and $g_{o}(\rho)=  \int_{\rho _{v}}^{\rho }\mu
_{o}(\rho)\,d\rho\ $ is   null for  liquid and vapor  bulks of phase
equilibrium. The pressure $P$ is
\begin{equation}
P(\rho)=\rho \, \mu _{o}(\rho)-g_{o}(\rho)\ +P_{o}
.\label{therm.pressure}
\end{equation}
Thanks to Eq. (\ref{equilibrium2a}), we obtain in the fluid
\emph{and not only in the fluid interlayer, }
\begin{equation*}
\mu _{o}(\rho)-\lambda \Delta \rho =\mu _{{o}}(\rho _{b})
,\label{equilibrium2b}
\end{equation*}
where $\mu _{{o}}(\rho _{b})$ is the chemical potential value of  a
liquid \emph{mother} bulk  of density  $\rho _{b}$
 such that $\mu _{{o}}(\rho _{b})= \mu
_{{o}}(\rho_{v_{b}})$, where $\rho_{v_{b}}$ is the density of the
vapor\emph{ mother} bulk bounding the layer.  We must emphasis that
$P(\rho _{b})$ and $P(\rho_{v_{b}})$ are unequal as for drop or
bubble  bulk pressures. Density $\rho_b$ is not a fluid density in
the interlayer but density in the liquid bulk
  from which the interface layer can extend (this is the reason why Derjaguin used the term \emph{mother liquid}  \cite{Derjaguin}, \emph{page 32}).
In the interlayer
 \begin{equation}
\lambda\,\frac{d^2\rho}{dz^2} = \mu_{b}(\rho), \quad {\rm with}\quad
\mu_{b}(\rho) = \mu_o(\rho)-\mu_o(\rho _{b}) \label{equilibrium2d}
\end{equation}
\subsection{Boundary conditions}
Condition  at the solid wall $(S)$ is associated with  Eq.
(\ref{surface energy})  \cite{Gouin1}  :
\begin{equation}
\lambda \left(\frac{d\rho }{dn}\right)_{|_S}+\phi ^{\prime
}(\rho)_{|_S}\ =0, \label{cl1}
\end{equation}
where $n$ is the external normal direction to the fluid;  Eq.
(\ref{cl1}) yields
\begin{equation*}
\lambda \left(\frac{d\rho }{dz}\right)_{|_{z=0}}=-\gamma _{1}+\gamma
_{2\ }\rho_{|_{z=0}} . \label{BC1}
\end{equation*}
Condition at the liquid-vapor interface  $(\Sigma)$ is associated
with Eq. (\ref{cl2}):
\begin{equation}
\lambda \left(\frac{d\rho }{dz}\right)_{|_{z=h}}=-\gamma _{4}\
\rho_{|_{z=h}}\,.  \label{BC2}
\end{equation}
Eq. (\ref{BC2}) defines the film thickness by introducing a
reference point inside the liquid-vapor interface bordering the
liquid interlayer with a convenient density at
$z=h$\cite{Gavrilyuk}.\newline We must also add the classical
surface conditions on the stress vector associated  with the total
stress tensor $\sigma+\sigma_v$ to these conditions on density.

\section{The disjoining pressure for  horizontal liquid films}

We consider  fluids and solids  \emph{at a given temperature}
$\theta$.   The hydrostatic pressure in a thin liquid interlayer
included between a solid wall and a vapor bulk differs from the
pressure in the contiguous liquid phase. At equilibrium, the
additional pressure interlayer is called the \emph{disjoining
pressure} \cite{Derjaguin}. The measure of a disjoining pressure is
either the additional pressure on the surface or the drop in the
pressure within the \emph{mother bulks} that produce the interlayer.
The disjoining pressure is equal to the difference between
  the pressure $P_{{v_b}}$ on the interfacial surface (pressure of the  vapor mother bulk  of density $\rho_{v_b}$) and the
  pressure $P_b$ in the  liquid mother bulk  (density $\rho_b$)
   from which the interlayer extends :
\begin{equation*}
\Pi(h) = P_{{v_b}}-P_b .\label{disjoiningpressure}
\end{equation*}
If $  g_{b}(\rho) =
g_o(\rho)-g_o(\rho_b)-\mu_{o}(\rho_{b})(\rho-\rho_{b})$ denotes the
primitive of  $\mu _{b}(\rho)$ null for $\rho _{b}$, we get  from
Eq. (\ref{therm.pressure})
\begin{equation}
\Pi (\rho _{b}) = -g_{b}(\rho_{v_{b}}), \label{disjoining}
\end{equation}
and an integration of Eq. (\ref{equilibrium2d}) yields
\begin{equation}
\frac{\lambda }{2}\,\left(\frac{d\rho }{dz}\right)^2=g_{b}(\rho)+\Pi
(\rho _{b}). \label{equilibrium2e}
\end{equation}
The reference chemical potential linearized near $\rho_l$
(\emph{respectively} $\rho_v$) is $\ \mu _{o}(\rho)=\displaystyle
\frac{c_{l}^{2}}{\rho _{l}}(\rho -\rho_{l})\ $ (\emph{respectively}
$\ \mu _{o}(\rho)=\displaystyle  \frac{c_{v}^{2}}{\rho _{v}}(\rho
-\rho_{v})\ $) where $c_l$ (\emph{respectively} $c_v$)  is the
isothermal sound velocity in liquid bulk $\rho_l$
(\emph{respectively vapor bulk} $\rho_v$) at temperature $\theta$
\cite{espanet}. In the liquid and vapor parts of the liquid-vapor
film,
 Eq. (\ref{equilibrium2d})  yields
\begin{equation*}
\lambda \frac{d^{2}\rho }{dz^{2}} =\frac{c_{l}^{2}}{\rho _{l}}(\rho
-\rho _{b})\quad {\rm ( liquid)}\quad {\rm and} \quad \lambda
\frac{d^{2}\rho }{dz^{2}}=\frac{c_{v}^{2}}{\rho _{v}}(\rho -\rho
_{v_{b}}) \quad {\rm ( vapor)}.\label{liquidensity}
\end{equation*}
The values of $\mu _{o}(\rho)$ are equal for the mother densities
$\rho _{v_{b}}$ and $\rho _{{b}}$,
\begin{equation*}
\frac{c_{l}^{2}}{\rho _{l}}(\rho _{b}-\rho _{l}) =\mu _{o}(\rho
_{b})=\mu _{o}(\rho _{v_{b}}) =\frac{c_{v}^{2}}{\rho _{v}}(\rho
_{v_{b}}-\rho _{v}), \ \  \rm and\ consequently,\label{densities1}
\end{equation*}
\begin{equation*}
\rho _{v_{b}}=\rho _{v}\left(
1+\frac{c_{l}^{2}}{c_{v}^{2}}\frac{(\rho _{b}-\rho _{l})}{\rho
_{l}}\right) .  \label{densities2}
\end{equation*}
In  liquid and vapor parts of the liquid-vapor interlayer we have,
\begin{equation*}
g_{o}(\rho)=\frac{c_{l}^{2}}{2\rho _{l}}(\rho -\rho _{l})^{2}\ \ \
{\rm ( liquid)}\quad {\rm and} \quad
g_{o}(\rho)=\frac{c_{v}^{2}}{2\rho _{v}}(\rho -\rho _{v})^{2}\ \ \
{\rm ( vapor)}.
\end{equation*}
From definition of $g_b(\rho)$ and Eq. (\ref{disjoining}) we deduce
the disjoining pressure
\begin{equation}
\Pi (\rho _{b })=\frac{c_{l}^{2}}{ 2\rho _{l}}(\rho _{l}-\rho _{b })%
\left[ \rho _{l}+\rho _{b}-\rho _{v}\left( 2+\frac{c_{l}^{2}}{c_{v}^{2}}%
\frac{(\rho _{b}-\rho _{l})}{\rho _{l}}\right) \right] .
\label{disjoining pressure2}
\end{equation}
Due to\ $\ \displaystyle\rho _{v}\left( 2+\frac{%
c_{l}^{2}}{c_{v}^{2}}\frac{(\rho _{b}-\rho _{l})}{\rho _{l}}\right)
\ll \rho _{l}+\rho _{b}$, we get $\displaystyle\
 \Pi
(\rho _{b})\approx\frac{c_{l}^{2}}{2\rho _{l}}(\rho _{l}^{2}-\rho
_{b}^{2}) . $ Now, we consider a film   of thickness  $h$; the
density profile in the liquid part of the liquid-vapor film is
solution of system :
\begin{equation*}
\left\{
\begin{array}{c}\quad \quad\quad \quad \quad\quad \quad\quad \quad\quad\quad\quad \quad
\displaystyle\lambda \frac{d^{2}\rho }{dz^{2}}=\frac{c_{l}^{2}}{\rho
_{l}}
(\rho -\rho _{b})\quad \quad\quad \quad \quad\quad \quad \quad\   (S1) \\
\quad  {\rm with}\quad \displaystyle\lambda \frac{d\rho
}{dz}_{\left| _{z=0}\right. }=-\gamma _{1}+\gamma _{2\ }\rho
_{\left| _{z=0}\right. }\quad {\rm and}\quad \displaystyle\lambda
\frac{d\rho }{dz}_{\left| _{z=h}\right. }=-\gamma _{4}\ \rho
_{\left| _{z=h}\right. }.
\end{array}
\right.
\end{equation*}
Quantity $\tau $ is defined such that $\displaystyle \tau
 ={c_{l}}/{\sqrt{\lambda \rho _{l}}}\ , $ where $1/\tau $ is a
reference length   and    $ \gamma _{3}=\lambda \tau$. Solution of
system $(S1)$ is
\begin{equation}
\rho =\rho _{b}+\rho _{1}\,e^{-\tau z}+\rho _{2}\,e^{\tau z}
\label{profil}
\end{equation}
where boundary conditions at $z=0$ and $h$ yield the values of $\rho
_{1}$ and $\rho _{2}$ :
\begin{equation*}
\left\{
\begin{array}{c}
\quad \quad\quad\quad \quad \quad\quad \quad \quad\quad (\gamma _{2}+\gamma _{3})\rho _{1}+(\gamma _{2}-\gamma _{3})\rho
_{2}=\gamma
_{1}-\gamma _{2}\rho _{b} \quad \quad\quad\ (S2)\\
\ \ \ \quad \quad -e^{-h\tau }(\gamma _{3}-\gamma _{4})\rho
_{1}+e^{h\tau }(\gamma _{3}+\gamma _{4})\rho _{2}=-\gamma _{4}\rho
_{b}.
\end{array}
\right. \
\end{equation*}
The liquid density profile is a consequence of Eq. ({\ref{profil})
when   $z$ $\in \left[ 0,h\right]$.  Taking Eq. ({\ref{profil}) into
account  in Eq. (\ref {equilibrium2e}) and $g_{b}(\rho)=
(c_l^2/2\,\rho_l)(\rho-\rho_b)^2$ in linearized form for the liquid
part of the   interlayer,  we get
\begin{equation}
\Pi (\rho _{b})=-\frac{2\,c_{l}^{2}}{\rho _{l}}\,\rho _{1}\, \rho
_{2}.  \label{Derjaguine}
\end{equation}
By identification of expressions  (\ref{disjoining pressure2}),
(\ref {Derjaguine}) and using $(S2)$, we get a relation between $h$ and $\rho _{b}$.
We denote finally the disjoining pressure by
 $\Pi (h)$.
\newline
Due to the fact that $\rho _{b}\simeq \rho _{l}$ \cite{Derjaguin},
the disjoining pressure reduces to
\begin{eqnarray*}
\Pi (h) &=&\frac{2\,c_{l}^{2}}{\rho _{l}}\left[ (\gamma _{1}-\gamma
_{2}\rho _{l})(\gamma _{3}+\gamma _{4})e^{h \tau }+(\gamma
_{2}-\gamma
_{3})\gamma _{4}\rho _{l}\right] \times  \notag \\
&&\frac{\left[ (\gamma _{2}+\gamma _{3})\gamma _{4}\rho _{l}-(\gamma
_{1}-\gamma _{2}\rho _{l})(\gamma _{3}-\gamma _{4})e^{-h \tau }\right] }{%
\left[ (\gamma _{2}+\gamma _{3})(\gamma _{3}+\gamma _{4})e^{h \tau
}+(\gamma_{3}-\gamma _{4})(\gamma _{2}-\gamma _{3})e^{-h \tau
}\right] ^{2}} .\label{Derjaguine bis}
\end{eqnarray*}
\emph{Let us notice  an important property}   of  mixture of   van
der Waals fluid and  perfect gas where the total pressure is the
sum of partial pressures of components \cite{espanet}: at
equilibrium, the partial pressure of the perfect gas is constant
through the liquid-vapor-gas interlayer -where  the perfect gas is
dissolved in the liquid. The  disjoining pressure of the mixture is
the same than for a single van der Waals fluid and calculations and
results are identical to those previously obtained.

\section{Motions along a liquid nanolayer}

When the liquid layer thickness  is small with respect to transverse
dimensions of the wall, it is possible to simplify  the
Navier-Stokes equation which governs the flow of a classical viscous
fluid  in the approximation of lubrication
\cite{batchelor}. When  $h\ll L$, where    $L$ is the wall
transversal characteristic size,   \emph{i)} the velocity component
along the wall is large with respect to the normal velocity
component which  can be neglected ;  \emph{ii)} the
velocity vector varies mainly along the  direction orthogonal to the wall and
 it is possible to neglect   velocity derivatives with
respect to coordinates along  the wall compared to the normal
derivative ; \emph{iii)} the pressure is constant in the direction
normal to the wall. It is possible to neglect the inertial term when
 $Re\ll L/h $ ($Re$ is the Reynolds  number of the flow).
Equation of Navier-Stokes is not valid in a liquid nanolayer because
the fluid is strongly inhomogeneous and the elastic stress tensor is
not scalar. However, it is possible to adapt the approximation of
lubrication for  viscous  flows in a liquid nanolayer.
 We are in the case of long wave
approximation:   $\  \epsilon =h/L\ll 1$.   We denote the velocity by
$\mathbf{V}=(u,v,w)$ where $(u,v)$ are the tangential components. In
the approximation of lubrication we have : $ \ \
e=\mathrm{sup}\left( \left| w/u\right| ,\left| w/v\right| \right)
\ll 1$. The main parts of  terms associated with second derivatives
 of  liquid velocity components correspond to   $ {\partial ^{2}u}/{\partial z^{2}}$ and ${\partial ^{2}v}/{
\partial z^{2}}$.
The density is constant along each stream line  ($\overset{
\mathbf{\centerdot}}{\rho }=0\Longleftrightarrow div \mathbf{V}=0$)
and iso-density surfaces
 contain the
trajectories. Then, $ {\partial u}/{\partial x} , {
\partial v}/{\partial y}$   and  $ {\partial w}/{\partial z}$
 have the same order of magnitude and    $\epsilon \sim e$.
 As in
Rocard model,  we assume  that the kinematic viscosity coefficient
$\nu= \kappa_2/\rho$ depends only on the temperature \cite{Rocard}.
In motion  equation, the viscosity term is $\
({1}/{\rho})\text{ div }\mathbf{\sigma }_{v}=2 \nu \ \left[\  \text{div }%
D\,+\, D\text{ grad  \{\,Ln}\,  (2\,\kappa_2)\}
 \ \right];
$
 $D$ grad\{Ln ($ 2\,\kappa_2$)\} is negligible with
respect to div $D$. In both lubrication and long wave approximations
the liquid nanolayer motion  verifies
\begin{equation}
{\mathbf{\Gamma }}+\text{grad}[\, \mu _{o}(\rho)-\lambda\,\Delta
\rho\,]=\nu \,\Delta {\mathbf{V}}\quad  {\rm with}\quad \Delta
{\mathbf{V}}\simeq
\begin{bmatrix} \displaystyle\;\frac{\partial ^{2}u}{\partial
z^{2}}, \displaystyle\;\frac{\partial ^{2}v}{\partial z^{2}}, 0
\end{bmatrix}\label{motion}
\end{equation}
In   approximation of lubrication, the inertial term is neglected
and Eq. ({\ref{motion}) separates into  tangential and normal
components to the solid wall. As in equilibrium, the normal
component of Eq. (\ref{motion}) is
\begin{equation*}
\frac{\partial }{\partial z}\left[ \ \mu _{o}(\rho )-\lambda
\,\Delta \rho\, \right] =0 \quad\Rightarrow\quad \mu _{o}(\rho
)-\lambda \,\Delta \rho =\mu _{o}(\rho _{b}).
\end{equation*}
To each value  $\rho _{b}$ (different of liquid bulk density value
$\rho_l$ of the plane interface  at equilibrium) is associated a
liquid nanolayer thickness $h$. We can write $ \mu _{o}(\rho
)-\lambda \,\Delta \rho =\eta (h), $
 where $\eta$ is such that $\eta (h) = \mu _{o}(\rho _{b})$.
For one-dimensional motions colinear
 to the solid wall (direction
 ${\mathbf{i_o}}$ and velocity  $u\,{\mathbf{i_o}}$),
 the tangential component of Eq.
(\ref{motion})
  yields :
\begin{equation}
\text{grad{\ }}\mu _{o}(\rho _{b})=\nu \frac{\partial ^{2}u}{\partial z^{2}}%
\ {\mathbf{i_{o}}}\ \ \ \Longleftrightarrow \ \ \ \frac{\partial \mu _{o}}{%
\partial \rho _{b}}\ \frac{\partial \rho _{b}}{\partial x}=\ \nu \frac{%
\partial ^{2}u}{\partial z^{2}}.  \label{viscosity}
\end{equation}
A liquid can slip on a solid wall only at a molecular level
\cite{Chuarev}. The sizes of  solid walls are    several orders of
magnitude higher than slipping distances which are negligible and
kinematic condition  at solid walls is the   adherence condition
$({z=0 \;\Rightarrow \;u=0})$. From the continuity of fluid
tangential stresses through a liquid-vapor interface of molecular
size and assuming that  vapor viscosity stresses  are negligible, we
obtain $\displaystyle( z=h, \;\Rightarrow\; \frac{\partial
u}{\partial z}=0)$. Consequently  Eq. (\ref{viscosity}) implies $\
\displaystyle \nu \,u=\frac{\partial \mu _{o}}{\partial \rho _{b}}\
\frac{\partial \rho _{b}}{\partial x}\ \left(
\frac{1}{2}\,z^{2}-h\,z\right) . $ The mean spatial velocity
$\overline{u}$ of the liquid in the nanolayer is  $ {\
\overline{u}=\frac{1}{h}\int_{o}^{h}u\ dz} $; previous computations
yield $\displaystyle \ \nu \
{\mathbf{\overline{u}}}=-\frac{h^{2}}{3}\ \text{grad}\ \mu _{o}(\rho
_{b}) \quad {\rm with}\quad {\mathbf{\overline{u}}} = \overline{u}\
{\mathbf{i_o}}. $ Let us remark that
\begin{equation*}
\frac{\partial \mu _{o}(\rho _{b})}{\partial x}=\frac{\partial \mu
_{o}}{\partial \rho_b }\,\frac{\partial \rho _{b}}{\partial
h}\,\frac{\partial h}{\partial x}\equiv \frac{1}{\rho
_{b}}\,\frac{\partial
P(\rho _{b})}{\partial \rho }\,\frac{\partial \rho _{b}}{\partial h}%
\,\frac{\partial h}{\partial x}.
\end{equation*}
The pressure  $P_{v_{b}}$ in the vapor bulk is constant along flow
motions and  $\Pi (h)=P_{v_{b}}-P(\rho _{b})$; consequently, we get
$\displaystyle
\frac{\partial \mu _{o}(\rho _{b})}{\partial x}=-\frac{1}{\rho _{b}}\ \frac{%
\partial \Pi (h)}{\partial h}\ \frac{\partial h}{\partial x}
$\ \ and
\begin{equation}
\chi_b \ {\mathbf{\overline{u}}}=\frac{h^{2}}{3}\ \text{grad}\ \Pi
(h).\label{variation potentiel chimique}
\end{equation}
where $\chi_{b}=\rho _{b}\nu \;$ is the liquid kinetic
viscosity.\newline Eq. (\ref{variation potentiel chimique}) yields
the mean spatial velocity of the isothermal liquid nanolayer as a
function of the disjoining pressure gradient. Like as the disjoining
pressure depends on the nanolayer thickness, the mean flow velocity
is a function of thickness variations along the flow.\newline
  Taking into account that in the liquid nanolayer
$\rho\simeq\rho_b$, then
\begin{equation*}
\left(\int_{o}^{h}\rho \ dz\right)\, { \mathbf{\overline{u}}}\simeq
\int_{o}^{h}\rho\,\mathbf{u}\ dz
\end{equation*}
and the mean spatial velocity  corresponds also to the mean velocity
with respect to the mass density.\newline In shallow water
approximation, the equation of continuity yields
\begin{equation*} \frac{\partial}{\partial t}\left({\int_{o}^{h}\rho \
dz}\right) + {\rm div}\left\{\left({\int_{o}^{h}\rho \ dz}\right){
\mathbf{\overline{u}}}\right\} = 0
\end{equation*}
 and we obtain Eq. (14) in ref.
(\cite{Nikolaiev}) associated with $h-$perturbations :
\begin{equation*}\frac{\partial h}{\partial t} + h \ {\rm div} {
\mathbf{\overline{u}}}=0 .
\end{equation*}
Thanks to Eq. (\ref{variation potentiel chimique})   an equation for
$h-$perturbations is :
\begin{equation}
\frac{\partial h}{\partial t}
+\frac{h}{3\,\chi_b}\,\frac{\partial}{\partial x}\left(h^2
\frac{\partial}{\partial x} \Pi(h)\right) =0.\label{lubrication}
\end{equation}
Eq. (\ref{lubrication}) is an \emph{equation of  diffusion} in
parabolic structure with a good sign of diffusion coefficient
associated with stability when $\displaystyle\frac{\partial
\Pi(h)}{\partial h}<0$ \cite{Derjaguin}.
\section*{Acknowledgments}This paper   supported by
PRIN 2005 (Nonlinear Propagation and Stability in Thermodynamical Processes of Continuous Media) is dedicated to Prof. Tommaso Ruggeri.
\bibliographystyle{ws-procs9x6}

\begin{thebibliography}{99}

\bibitem{Widom}  B. Widom,
\emph{Physica A,} \textbf{263}, 500  (1999).

\bibitem{chernov1}  A.A. Chernov,  L.V. Mikheev, \emph{Phys. Rev. Lett},  \textbf{60}, 2488   (1988).

\bibitem{Evans1}  R. Evans, \emph{Adv. Phys.} \textbf{28}, 143  (1979).

\bibitem{Fisher} M.E. Fisher,  A.J. Jin, \emph{Phys. Rev. B,} \textbf{44}, 1430  (1991).

\bibitem{rowlinson}  J.S.  Rowlinson,  B. Widom, \emph{Molecular theory of
capillarity}, (Clarendon Press, Oxford, 1984).

\bibitem{Bhushan}  \emph{Springer handbook of nanotechnology,} Ed. B. Bhushan
(Berlin, 2004).

\bibitem{Fisher1} H. Nakanishi,  M.E. Fisher,   \emph{Phys. Rev. Lett.} \textbf{49}, 1565  (1982).

\bibitem{Cahn0}   J.W. Cahn,  \emph{J. Chem. Phys. }\textbf{66},
3667   (1977).

\bibitem{gouin}   H. Gouin, \emph{J. Phys. Chem. B} \textbf{102}, 1212
(1998); arXiv:0801.4481.

\bibitem{Rocard}  Y. Rocard, \emph{Thermodynamique}, (Masson, Paris, 1967).

\bibitem{Israel}  J. Israelachvili, \emph{Intermolecular  and surface forces},
(Academic Press, New York, 1992).

\bibitem{degennes}  P.G. de Gennes, \emph{Rev.
Mod. Phys.} \textbf{57},  827  (1985).

\bibitem{Gavrilyuk} S. Gavrilyuk, I. Akhatov, \emph{Phys. Rev. E.}, \textbf{73}, 021604 (2006).

\bibitem{gouin4}   H. Gouin, \emph{Physicochemical Hydrodynamics, B Physics}
\textbf{174},  667  (1987).

\bibitem{Gouin1}   H. Gouin, W. Kosi\'{n}ski, \emph{Arch. Mech.} \textbf{50}, 907
(1998); arXiv:0802.1995.

\bibitem{Gouin05} H. Gouin, T. Ruggeri, {\it Eur. J. Mech. B/fluids}, \textbf{24},
596 (2005); arXiv:0801.2096.

\bibitem{Derjaguin}  B.V. Derjaguin,  N.V. Churaev,  V.M. Muller, \emph{ Surfaces
forces,}  (Plenum Press, New York, 1987).

\bibitem{espanet}   H. Gouin, L. Espanet,
\emph{C. R. Acad. Sci. Paris}  {\textbf{328}, IIb,}  {151 } (2000);
arXiv:0807.5023.

\bibitem{batchelor}  G. K.  Batchelor, \emph{An Introduction to Fluid Dynamics,}
(Cambridge University Press, 1967).

\bibitem{Chuarev}  N.V. Churaev,   \emph{Colloid. J.} \textbf{58}, 681 (1996).

\bibitem{Nikolaiev}  V.S. Nikolayev,  S.L. Gavrilyuk,  H. Gouin,  \emph{J. Coll. Interf. Sci.} \textbf{302}, 605
(2006); arXiv:0802.2479.
\end{thebibliography}

\end{document}